\documentstyle[12pt,a4]{article}
\begin{document}
  \title{A spin-shift relation of the Ruijsenaars-Schneider
          Hamiltonian of type $C_n$}
 \author{Kazuyuki OSHIMA \\
     Graduate School of Mathematics,
       Nagoya University \\
     Chikusa-ku, Nagoya 464-8602, Japan \\
     e-mail: ooshima@math.nagoya-u.ac.jp}
\maketitle

\begin{abstract}
We construct a spin-shift relation of the trigonometric 
Ruijsenaars-Schneider Hamiltonian of type $C_n$. 
This is a successive study of the previous papar \cite{Oshima}.
\end{abstract}

\section{Introduction}
Ruijsenaars-Schneider systems describe one-dimensional $n$-particle
system with pairwise interaction. In \cite{BBB}, the authors derived
the Hamiltonian for the trigonometric one-particle Ruijsenaars-
Schneider system from the Gervais-Neveu-Felder equation. 
We would like to recall their argument in brief.
The Cervais-Neveu-Felder equation has its $L$-invariant form:
\begin{equation}
R_{12}(x q^{-H_{3}/2})L_{13}(x)L_{23}(x)=
 L_{23}(x)L_{13}(x)R_{12}(x q^{H_{3}/2}) 
\end{equation}
with a subscript $3$ denote the quantum space.
Let us represent $L_{13}(x)$ in the tensor product of the representation
$\rho^{\left(\frac{1}{2}\right)} \otimes \rho^{(j)}$,
where $\rho^{(j)}$ is the spin-$j$ representation of $U_{q}(sl_{2})$.
Taking the trace on the first space, and still restricting this 
operator to the space of zero-weight vectors, one obtains
\begin{equation}
 H_{j}= q^{-x \frac{\partial}{\partial x}}+
  q^{-x \frac{\partial}{\partial x}}\left( 1-
  \frac{(q^{j}-q^{-j})(q^{j+1}-q^{-j-1})}{(x-x^{-1})(q^{-1}x-q x^{-1})}
 \right)
\end{equation}
when $j$ is an integer. This is the trigonometric one-particle Ruijsenaars-
Schneider Hamiltonian. 
That is the reason we call $j$ the spin of the model insted of the
coupling constant of the model.
The integrability of the system generated by the Hamiltonian obtained
here was solved by using the existence of a spin-shift
operator $D_{j}$ which satisfies
\begin{equation}
 H_{j}D_{j}=D_{j}H_{j-1}.
\end{equation} 

In recent work \cite{Oshima}, we constructed the spin-shift operator
for the trigonometric multi-particle Ruijsenaars-Schneider system 
in terms of the Dunkl-Cherednik operators.

The purpose of the present paper is to find results analogous to
those of \cite{Oshima} for the root system $C_n$. As a consequence
we obtain the spin-shift operator in the language of Weyl group.

This Letter is arranged as follows. In Section 2, we review
the basic facts of the Dunkl-Cherednik operators and the Ruijsenaars-
Schneider Hamiltonian. Following the similar process as in \cite{Oshima}
leads us to obtain the spin-shift relation (Proposition 2) and the explicit
form of the spin-shift operator for trigonometric Ruijsenaars-
Schneider Hamiltonian of type $C_n$ (Theorem 1) in Section 3.

\section{Dunkl-Cherednik operators}

In this section we define the Dunkl-Cherednik operators, which are the key
tool
for our paper. For details, see references (\cite{C}-\cite{KJr}).

Let $V$ be an $n$-dimensional real vector space with the basis
$\{ \epsilon_1, \dots, \epsilon_n \}$, and a positive definite 
symmetric bilinear form
$(\cdot, \cdot): V \times V \to {\bf R}$ defined by 
$(\epsilon_i,\epsilon_j)=\delta_{ij}$.
Let $R=\{ \pm \epsilon_i \pm \epsilon_j \, (1 \le i < j \le n), \,
          \pm 2 \epsilon_i \, (1 \le i \le n)\} $
be the root system of type $C_n$,
$R^{+}= \{\epsilon_i \pm \epsilon_j \, (1 \le i < j \le n), \,
          2 \epsilon_i \, (1 \le i \le n) \}$
the set of positive roots,
$\Pi = \{ \alpha_i =\epsilon_i-\epsilon_{i+1} \, (1 \le i \le n-1), 
 \alpha_{n}=2 \epsilon_n \}$
the set of simple roots.
For every root $\alpha \in R$, define the coroot
$\alpha^{\vee} := 2 \alpha/(\alpha, \alpha)$.
Denote by $P=\{\lambda \in V | (\lambda, \alpha_{i}^{\vee}) \in {\bf Z} \}$
the weight lattice. It has a natural basis of fundamental weights $\omega_i$
determined by $(\alpha_{j}^{\vee},\omega_{i})=\delta_{ji}$.
In similar way we define coweight lattice 
$P^{\vee}=\{ \lambda^{\vee} \in V | (\lambda^{\vee}, \alpha_i) \in {\bf Z} \}$
and fundamental coweight $\omega_{i}^{\vee}$ determined by 
$(\alpha_j, \omega_{i}^{\vee})=\delta_{ji}$. 
As usual we define the highest root $\theta \in R$ by
$\theta - \alpha \in \oplus \, {\bf Z}_{+} \alpha_{i}$
for all $\alpha \in R$.

Let 
$\Lambda_n = {\bf Q}(q)[X^{\epsilon_1}, \dots, X^{\epsilon_n}]$ 
be a polynomial ring.
For every $ \alpha \in R$ denote $s_{\alpha}$ for the reflection with
respect to $\alpha$ determined by $s_{\alpha}(v)=v-(v,\alpha)\alpha^{\vee}$.
The Weyl group $W=W_{C_n}= \langle s_1, \dots, s_{n-1}, s_n \rangle $ 
acts on $\Lambda_n$ as $w. X^{\alpha}=X^{w(\alpha)}$.

The set of affine roots is
$\widetilde{R}=\{ \alpha +m \delta |  \alpha \in R, m \in {\bf Z} \}$, 
where $\delta$ denotes the constant function $1$ on $V$.
The simple roots  are $a_0=-\theta + \delta$ and $a_i=\alpha_i \in R$.
We use the same symbols $s_i \, (0 \le i \le n)$ to represent the generator
for the corresponding affine Weyl group 
$\widetilde{W}= \langle s_0, s_1, \dots, s_n \rangle $.
We note that $s_0=\tau(\theta^{\vee})s_{\theta}$, where $\tau(\xi^{\vee})$
is defined by $ \tau(\xi^{\vee})X^{\mu}=q^{2(\xi^{\vee},\mu)}X^{\mu}$.
We define the length $\ell(\lambda^{\vee})$ of $\lambda^{\vee} \in P^{\vee}$
as the length $l$ of the reduced expression
\begin{equation}
 \tau(\lambda^{\vee}) = s_{i_1} \cdots s_{i_{l}}.
\end{equation}
We write $ \lambda^{\vee} \prec \mu{\vee}$ if
$\ell(\lambda^{\vee}) < \ell(\mu^{\vee})$.
Suppose for every $ \alpha \in \widetilde{R}$ we have a variable 
$t_{\alpha}$ such that $t_{\alpha}=t_{w(\alpha)}$ for every 
$ w \in \widetilde{W}$. 
Let ${\bf Q}_{t}= {\bf Q}(t_{\alpha}) $ be the 
field of rational functions in $t_{\alpha}$.
We now introduce the operators
\begin{equation}
 T_{i} = t_{\alpha_{i}}s_{i} + (t_{\alpha_{i}}-{t_{\alpha_{i}}}^{-1})
 \frac{1}{X^{-\alpha_{i}}-1}(s_{i}-1) \quad (i=0, \dots, n).
\end{equation}
They satisfy the following relations
\begin{eqnarray}
 &&(T_{i}-t_{\alpha_{i}})(T_{i}+{t_{\alpha_{i}}}^{-1})=0 \quad
 (i=0,\dots, n),  \label{aha1}\\
 && T_{i}T_{j}=T_{j}T_{i} \quad (|i-j| \ge 2),  \label{aha2}\\
 && T_{i}T_{i+1}T_{i}=T_{i+1}T_{i}T_{i+1} \quad 
 (i=1, \dots , n-2), \label{aha3}\\
 && T_{i}T_{i+1}T_{i}T_{i+1}=T_{i+1}T_{i}T_{i+1}T_{i} \quad
 (i=0,n-1) \label{aha4}.
\end{eqnarray}
This means that they realize the representation of the affine Hecke algebra
$H(\widetilde{W})$ of $\widetilde{W}$ on 
$\Lambda_{n,t}={\bf Q}_{t}(q)[X^{\epsilon_1},\dots, X^{\epsilon_n}]$.

The Dunkl-Cherednik operator $Y^{\omega_{i}^{\vee}} \enskip (i=1,\dots,n)$
are now defined in terms of the operators $T_i \enskip (i=0, \dots, n)$ by
the expression
\begin{equation}
  Y^{\omega_{i}^{\vee}}:=(T_{0}T_{1} \cdots T_{n-1}T_{n}T_{n-1} \cdots
T_{i})^{i}.
\end{equation}
Using the relations (\ref{aha2})-(\ref{aha4}), one can show that the
Dunkl-Cherednik operators 
$Y^{\omega_{1}^{\vee}}, \dots , Y^{\omega_{n}^{\vee}}$ commute with
each other.
Moreover they satisfy the following commutation relations with 
$T_{i}$.
\begin{equation}
 \left\{
\begin{array}{ll}
   T_i Y^{\omega^{\vee}} - Y^{s_{i}(\omega^{\vee})}T_i
        =(t_{\alpha_{i}}-t_{\alpha_{i}}^{-1})Y^{\omega^{\vee}}, &
          \mbox{if $(\omega^{\vee},\alpha_i)=1$,} \\
   T_i Y^{\omega^{\vee}} = Y^{\omega^{\vee}} T_i, &
          \mbox{otherwise.}
\end{array}\right.
\end{equation}

Set 
$\hat{R}=\{\alpha \in R\, | \, (\alpha , \omega_{n}^{\vee})=1 \}$.
Let $M(q, t_{\alpha})$ denote the Macdonald difference operator
\begin{equation}
 M(q,t_{\alpha})=
 \sum_{w \in W} \prod_{\alpha \in \hat{R}}
  \frac{t_{\alpha}X^{w(\alpha)}-{t_{\alpha}}^{-1}}{X^{w(\alpha)}-1}
  \tau(w(\omega_{n}^{\vee}))
\label{macd}
\end{equation}
It is known that the Macdonald difference operators are obtained 
in terms of the Dunkl-Cherednik operators.
\newtheorem{prop}{Proposition}
\begin{prop}\cite{KJr} 
\begin{equation}  \left. \sum_{w \in W}Y^{w(\omega_{n}^{\vee})}
  \right|_{\Lambda_{n,t}^{W}}= M(q, t_{\alpha})
   \label{prop1}
\end{equation}
Note that $O|_{\Lambda_{n,t}^{W}}$ means that the action of
the operator $O$ is restricted to the $W$-invariant space
$ \Lambda_{n,t}^{W}$.
\end{prop}

We will define the Ruijsenaars-Schneider Hamiltonian of type $C_n$
by conjugating the Macdonald difference operator (Equation (\ref{macd})) 
by a weight function as follows.

For this purpose we introduce the notation  
\begin{equation}
 (x;q^2)_{l}=\prod_{i=0}^{l-1}(1-x q^{2i}).
\end{equation}
Note that in terms of this notation the Macdonald difference operator
with $t_{\alpha}=q^{l_{\alpha}}$ can be written as
\begin{equation}
 M(q, q^{l_{\alpha}}) = \sum_{w \in W} \prod_{\alpha \in R^{+}}
 \frac{(X^{\alpha};q^2)_{l_{\alpha}}}{(X^{w(\alpha)};q^2)_{l_{\alpha}}}
 \tau(w(\omega_{n}^{\vee}))
\end{equation}
Let 
\begin{equation}
 \Delta^{+}:= \prod_{\alpha \in R^{+}} (X^{\alpha};q^2)_{l_{\alpha}}
\end{equation}
be the weight function we mentioned above. Consider an operator
\begin{equation}
 H_{(l_{\alpha})}:= \Delta^{+} M(q,q^{l_{\alpha}}) (\Delta^{+})^{-1},
\label{crs-hamiltonian}
\end{equation}
where 
$(l_{\alpha})=(l_{1},l_{2}) \in {\bf Z}_{\ge 0} \times {\bf Z}_{\ge 0}$.
The nonnegative integers $l_1$ and $l_2$ correspond to long root and
short root respectively. 
Now let
\begin{eqnarray*}
 \hat{R}_{w}^{+} &=& w(\hat{R}) \cap R^{+}, \\
 \hat{R}_{w}^{-} &=& -w(\hat{R}) \cap R^{+},
\end{eqnarray*}
then for $ \alpha \in R^{+}$
\[ \tau(w(\omega_{n}^{\vee}))(X^{\alpha};q^2)_{l_{\alpha}}= \left\{
 \begin{array}{rl}
  (q^2 X^{\alpha};q^2)_{l_{\alpha}}, & \quad 
   \mbox{for $ \alpha \in \hat{R}_{w}^{+}$} \\
  (q^{-2} X^{\alpha};q^2)_{l_{\alpha}}, & \quad
   \mbox{for $ \alpha \in \hat{R}_{w}^{-}$} \\
  (X^{\alpha};q^2)_{l_{\alpha}}, & \quad
   \mbox{otherwise.}
 \end{array} \right. \]
Therefore the Equation.(\ref{crs-hamiltonian}) is
\begin{equation}
 H_{(l_{\alpha})}= \sum_{w \in W} \prod_{\alpha \in R^{+}}
   \frac{q^{2 l_{\alpha}}X^{w(\alpha)}-1}{X^{w(\alpha)}-1}
  \prod_{\beta \in \hat{R}_{w}^{+}}
   \frac{X^{\beta}-1}{q^{2 l_{\beta}}X^{\beta}-1}
  \prod_{\gamma \in \hat{R}_{w}^{-}}
   \frac{q^{2 l_{\gamma}-2}X^{w(\gamma)}-1}{q^{-2}X^{w(\gamma)}-1}
 \tau(w(\omega_{n}^{\vee})) 
\label{CRS}
\end{equation}
We call the operator $H_{(l_{\alpha})}$
Ruijsenaars-Schneider Hamiltonian of type $C_n$ and  
$(l_{\alpha}) \in {\bf Z}_{\ge 0} \times {\bf Z}_{\ge 0}$ 
a spin of the model.

\section{Spin-Shift relation}

We begin with the equation which follows from the commutativity
of the Dunkl-Cherednik operators:
\begin{equation}
 \sum_{w \in W} Y^{w(\omega_{n}^{\vee})}
 \prod_{\alpha \in R^{+}}\left(t_{\alpha}^{-1} Y^{\frac{\alpha^{\vee}}{2}}
   -t_{\alpha}Y^{-\frac{\alpha^{\vee}}{2}} \right)=
 \prod_{\alpha \in R^{+}}\left(t_{\alpha}^{-1} Y^{\frac{\alpha^{\vee}}{2}}
   -t_{\alpha}Y^{-\frac{\alpha^{\vee}}{2}} \right)
 \sum_{w \in W} Y^{w(\omega_{n}^{\vee})}
 \label{start}
\end{equation}

\begin{prop}
Let $w_0$ be the longest element of Weyl group $W$. Multiplying
both sides of Equation (\ref{start}) by $w_0$ from the left, and
restricting the operators to $\Lambda_{n,t}^{W}$,
we obtain
\begin{eqnarray}
 && \left\{ \sum_{w \in W} \prod_{\alpha \in \hat{R}} 
  \frac{t_{\alpha}X^{w(\alpha)}-t_{\alpha}^{-1}}{X^{w(\alpha)}-1}
 \prod_{\beta \in \hat{R}_{w}^{-}}
  \frac{t_{\beta}X^{\beta}-t_{\beta}^{-1}}
        {t_{\beta}^{-1}X^{\beta}-t_{\beta}}
  \frac{q^{-2}t_{\beta}^{-1}X^{\beta}-t_{\beta}}
        {q^{-2}t_{\beta}X^{\beta}-t_{\beta}^{-1}} 
   \tau(w(\omega_{n}^{\vee})) \right\} \nonumber \\
 && \times 
  \left\{\prod_{\alpha \in R^{+}} 
   \frac{t_{\alpha}X^{w(\alpha)}-t_{\alpha}^{-1}}{X^{\alpha}-1}
  \right. \nonumber \\
 && \quad \times \left. \sum_{w \in W} {\rm sgn}(w) w \left( 
  \prod_{i=2}^{\infty} 
  \prod_{{\alpha \in R^{+}} \atop {ht(\alpha) \ge i}}
   \frac{t_{\alpha}q^{2(i-1)}X^{\alpha}-t_{\alpha}^{-1}}
         {q^{2(i-1)X^{\alpha}-1}} \tau(\rho^{\vee}) 
         + \sum_{\lambda^{\vee} \prec \rho^{\vee} }
           G_{\lambda^{\vee}}\tau(\lambda^{\vee})
         \right) \right\}  \nonumber \\
 &&= \left\{\prod_{\alpha \in R^{+}}
   \frac{t_{\alpha}X^{w(\alpha)}-t_{\alpha}^{-1}}{X^{\alpha}-1}
    \right. \nonumber \\
 && \quad \times  \left. \sum_{w \in W} {\rm sgn}(w) w \left(
  \prod_{i=2}^{\infty}
  \prod_{{\alpha \in R^{+}} \atop {ht(\alpha) \ge i}}
   \frac{t_{\alpha}q^{2(i-1)}X^{\alpha}-t_{\alpha}^{-1}}
         {q^{2(i-1)X^{\alpha}-1}} \tau(\rho^{\vee}) 
          + \sum_{\lambda^{\vee} \prec \rho^{\vee}}
           G_{\lambda^{\vee}}\tau(\lambda^{\vee})
         \right) \right\}  \nonumber \\
 && \times
  \left\{ \sum_{w \in W} \prod_{\alpha \in \hat{R}} 
   \frac{t_{\alpha}X^{w(\alpha)}-t_{\alpha}^{-1}}{X^{w(\alpha)}-1}
    \tau(w(\omega_{n}^{\vee})) \right\} 
  \label{prop}
\end{eqnarray}
where $G_{\lambda^{\vee}}$ is come meromorphic function in 
$\Lambda_{n,t}$.
\end{prop}

{\it proof}. \enskip
We can prove the proposition in the same manner as in \cite{Oshima}.
We will outline the proof.

Since in 
$\prod_{\alpha \in R^{+}} 
(t_{\alpha}^{-1}Y^{\frac{\alpha^{\vee}}{2}}-
 t_{\alpha}Y^{-\frac{\alpha^{\vee}}{2}})|_{\Lambda_{n,t}^{W}}$,
the term of $\tau(-\rho^{\vee})$ appears only from 
$Y^{-\rho^{\vee}}$ one can calculate the coefficient:
\begin{equation}
 \prod_{i=1}^{2n-1}\prod_{{\alpha \in R^{+}} \atop {ht(\alpha)\ge i}}
 \frac{t_{\alpha}-q^{-2(i-1)}t_{\alpha}^{-1}X^{\alpha}}
   {1-q^{-2(i-1)}X^{\alpha}}.
\label{coeff}
\end{equation}
Using the identity
\begin{eqnarray*}
 \lefteqn{(T_{i}+t_{\alpha_{i}}^{-1})\prod_{\alpha \in R^{+}}
  (t_{\alpha}^{-1}Y^{\frac{\alpha^{\vee}}{2}}
      -t_{\alpha}Y^{-\frac{\alpha^{\vee}}{2}})} \\
 &&=\frac{t_{\alpha_{i}}^{-1}Y^{-\frac{\alpha_{i}^{\vee}}{2}}
    -t_{\alpha_{i}} Y^{\frac{\alpha_{i}^{\vee}}{2}}}
    {t_{\alpha_{i}}^{-1}Y^{\frac{\alpha_{i}^{\vee}}{2}}
    -t_{\alpha_{i}} Y^{-\frac{\alpha_{i}^{\vee}}{2}}}
    \prod_{\alpha \in R^{+}}
      (t_{\alpha}^{-1}Y^{\frac{\alpha^{\vee}}{2}}
      -t_{\alpha}Y^{-\frac{\alpha^{\vee}}{2}})
      (T_{i}-t_{\alpha_{i}}),
\end{eqnarray*}
which can be shown by a direct calculation, we have
\begin{eqnarray}
 \lefteqn{\left. s_{i}\, \frac{1}{\prod_{\alpha \in R^{+}}
   \frac{t_{\alpha}-t_{\alpha}^{-1}X^{\alpha}}{1-X^{\alpha}}}
    \prod_{\alpha \in R^{+}}
      (t_{\alpha}^{-1}Y^{\frac{\alpha^{\vee}}{2}}
      -t_{\alpha}Y^{-\frac{\alpha^{\vee}}{2}}) 
         \right|_{\Lambda_{n,t}^{W}}} \nonumber \\
 &&=- \left. \frac{1}{\prod_{\alpha \in R^{+}}
   \frac{t_{\alpha}-t_{\alpha}^{-1}X^{\alpha}}{1-X^{\alpha}}}
    \prod_{\alpha \in R^{+}}
      (t_{\alpha}^{-1}Y^{\frac{\alpha^{\vee}}{2}}
      -t_{\alpha}Y^{-\frac{\alpha^{\vee}}{2}})\right|_{\Lambda_{n,t}^{W}}
 \label{antisym}
\end{eqnarray}
Thus this invarinaceness up to $(\pm 1)$ with Equation (\ref{coeff})
leads to
\begin{eqnarray}
  \lefteqn{\left. \prod_{\alpha \in R^{+}}
      (t_{\alpha}^{-1}Y^{\frac{\alpha^{\vee}}{2}}
      -t_{\alpha}Y^{-\frac{\alpha^{\vee}}{2}})
         \right|_{\Lambda_{n,t}^{W}}
  = \prod_{\alpha \in R^{+}} 
 \frac{t_{\alpha}-t_{\alpha}^{-1}X^{\alpha}}{1-X^{\alpha}}} \\
 && \times   \sum_{w \in W}{\rm sgn}(w)w \left(
   \prod_{i=2}^{2n-1}\prod_{{\alpha \in R^{+}} \atop {ht(\alpha)\ge i}}
 \frac{t_{\alpha}-q^{-2(i-1)}t_{\alpha}^{-1}X^{\alpha}}
   {1-q^{-2(i-1)}X^{\alpha}} \tau(-\rho^{\vee})
    +\sum_{\lambda^{\vee} \prec \rho^{\vee}} 
     G_{\lambda^{\vee}} \tau(-\lambda^{\vee}) \right), \nonumber
\end{eqnarray}
where $G_{\lambda^{\vee}}$ is some meromorphic function in $\Lambda_{n,t}$.
Since $\sum_{w \in W}Y^{w(\omega_{n}^{\vee})}|_{\Lambda_{n,t}^{W}}$
is Weyl invariant, the right hand side of Equation (\ref{prop})
is proved.

Next we step to prove the left hand side of Equation (\ref{prop}).
Let $\Omega^{\vee}$ be the Weyl orbit of $\omega_{n}^{\vee}$.
Because of the invarianceness up to $(\pm 1)$ (Equation (\ref{antisym})), 
we find that the equation
\begin{eqnarray}
 \lefteqn{ w_0 \left. \sum_{w \in W}Y^{w(\omega_{n}^{\vee})}
   \prod_{\alpha \in R^{+}}(t^{-1}Y^{\frac{\alpha^{\vee}}{2}}-
   t Y^{-\frac{\alpha^{\vee}}{2}}) \right|_{\Lambda_{n,t}^{W}}}\nonumber \\
 &&= w_0 \left. \prod_{\alpha \in R^{+}}
  (t^{-1}Y^{\frac{\alpha^{\vee}}{2}}- t Y^{-\frac{\alpha^{\vee}}{2}})
   \sum_{w \in W}Y^{w(\omega_{n}^{\vee})} \right|_{\Lambda_{n,t}^{W}}
\label{restrict}
\end{eqnarray}
turns to be
\begin{eqnarray}
 \lefteqn{\left\{ \sum_{\omega^{\vee} \in \Omega^{\vee}}C(\omega^{\vee})
   \tau(\omega^{\vee}) \right\}
   w_0 \left. \prod_{\alpha \in R^{+}}
    (t^{-1}Y^{\frac{\alpha^{\vee}}{2}}-
   t Y^{-\frac{\alpha^{\vee}}{2}}) \right|_{\Lambda_{n,t}^{W}}} \nonumber \\
  &&= w_0 \left. \prod_{\alpha \in R^{+}}
  (t^{-1}Y^{\frac{\alpha^{\vee}}{2}}- t Y^{-\frac{\alpha^{\vee}}{2}})
   \sum_{w \in W}Y^{w(\omega_{n}^{\vee})} \right|_{\Lambda_{n,t}^{W}}.
 \label{compare}
\end{eqnarray}
Now we will determine the coefficients $C(\omega^{\vee})$ as follows.
Note that we can write
\begin{eqnarray}
 \lefteqn{w_{0} \sum_{w \in W}
   Y^{w(\omega_{n}^{\vee})}} \label{top} \\ 
 && = \prod_{\alpha \in \hat{R}}
   \frac{t_{\alpha}X^{\alpha}-t_{\alpha}^{-1}}{X^{\alpha}-1}
   \tau(\omega_{n}^{\vee})+
 \sum_{\omega^{\vee}
  \in \Omega^{\vee} \setminus \{\omega_{n}^{\vee}\}}
 A(x_1,\dots,x_n)\tau(\omega^{\vee}) S,\nonumber
\end{eqnarray}
where $A(x_1,\dots,x_n)$ and $S$ are appropriate coefficients
and elements of Weyl group respectively.
Substituting Equation (\ref{top}) to (\ref{restrict}), 
multiplying $ w \in W$ to the both sides of (\ref{restrict}),
and comparing with (\ref{compare}),
we can observe that the term of $C(w(\omega_{n}^{\vee}))$ is
\begin{equation}
 \prod_{\alpha \in \hat{R}} 
 \frac{t_{\alpha}X^{w(\alpha)}-t_{\alpha}^{-1}}{X^{w(\alpha)}-1}
 \prod_{\beta \in \hat{R}_{w}^{-}} 
 \frac{t_{\beta}X^{\beta}-t_{\beta}^{-1}}{t_{\beta}^{-1}X^{\beta}-t_{\beta}}
 \frac{q^{-2}t_{\beta}^{-1}X^{\beta}-t_{\beta}}
    {q^{-2}t_{\beta}X^{\beta}-t_{\beta}^{-1}} \tau(w(\omega_{n}^{\vee})).
\end{equation}
Thus the left hand side of Equation (\ref{prop}) is proved. 
Hence the proof has completed. 
\hfill $\Box$ 
\medskip

From Equation (\ref{prop}), one can derive the spin-shift relation.
Put $t_{\alpha}=q^{l_{\alpha}}$, and conjugating by the weight function
$\Delta^{+}$, namely multiplying the both sides
of the above equation by $\Delta^{+}$ from the left and
$(\Delta^{+})^{-1}$ from the right, we obtain
\begin{eqnarray*}
 && \sum_{w \in W} \prod_{\alpha \in R^{+}}
   \frac{q^{2 (l_{\alpha}+1)}X^{w(\alpha)}-1}{q^2 X^{w(\alpha)}-1}
  \prod_{\beta \in \hat{R}_{w}^{+}}
   \frac{q^2 X^{\beta}-1}{q^{2 (l_{\beta}+1)}X^{\beta}-1}
  \prod_{\gamma \in \hat{R}_{w}^{-}}
   \frac{q^{2 l_{\gamma}}X^{w(\gamma)}-1}{X^{w(\gamma)}-1}
 \tau(w(\omega_{n}^{\vee})) \\
 && \times \Delta^{+} \hat{D}_{(l_{\alpha})} (\Delta^{+})^{-1}
   = \Delta^{+} \hat{D}_{(l_{\alpha})} (\Delta^{+})^{-1} \\
 &&   \times \sum_{w \in W} \prod_{\alpha \in R^{+}}
   \frac{q^{2 l_{\alpha}}X^{w(\alpha)}-1}{X^{w(\alpha)}-1}
  \prod_{\beta \in \hat{R}_{w}^{+}}
   \frac{X^{\beta}-1}{q^{2 l_{\beta}}X^{\beta}-1}
  \prod_{\gamma \in \hat{R}_{w}^{-}}
   \frac{q^{2 l_{\gamma}-2}X^{w(\gamma)}-1}{q^{-2}X^{w(\gamma)}-1}
 \tau(w(\omega_{n}^{\vee})).
\end{eqnarray*}
where
\begin{eqnarray}
 \lefteqn{\hat{D}_{(l_{\alpha})}=\prod_{\alpha \in R^{+}}
   \frac{t_{\alpha}X^{w(\alpha)}-t_{\alpha}^{-1}}{X^{\alpha}-1}}\\
  && \quad \times \sum_{w \in W} {\rm sgn}(w) w \left(
  \prod_{i=2}^{\infty}
  \prod_{{\alpha \in R^{+}} \atop {ht(\alpha) \ge i}}
   \frac{t_{\alpha}q^{2(i-1)}X^{\alpha}-t_{\alpha}^{-1}}
         {q^{2(i-1)X^{\alpha}-1}} \tau(\rho^{\vee}) 
   + \sum_{\lambda^{\vee} \prec \rho^{\vee}} 
   G_{\lambda^{\vee}}\tau(\lambda^{\vee}) \right). \nonumber
\end{eqnarray}

Now we calculate the explicit form of the spin-shift operator 
\begin{equation}
 D_{(l_{\alpha})}=\Delta^{+} \hat{D}_{(l_{\alpha})} (\Delta^{+})^{-1}.
\end{equation}
Since the action of $\tau(w(\rho^{\vee}))$ to 
$(X^{w(\alpha)};q^2)_{l_{\alpha}}$ is
\begin{equation}
  \tau(w(\rho^{\vee}))(X^{w(\alpha)}; q^2)_{l_{\alpha}} 
   = (q^{2 ht(\alpha)}X^{w(\alpha)};q^2)_{l_{\alpha}},
\end{equation}
we find
\begin{eqnarray*}
 \lefteqn{\tau(w(\rho^{\vee}))(\Delta^{+})^{-1}} \\
 &&=\prod_{\alpha \in \hat{R}_{w}^{+}}\prod_{i=0}^{l_{\alpha}-1}
  \frac{1}{1-q^{2(i+ht(\alpha))}X^{w(\alpha)}}
  \prod_{\alpha \in \hat{R}_{w}^{-}}\prod_{i=0}^{l_{\alpha}-1}
  \frac{1}{1-q^{2(i-ht(\alpha))}X^{-w(\alpha)}} \tau(w(\rho^{\vee})).
\end{eqnarray*} 
In conclusion we obtain the following result.

\newtheorem{th}{Theorem}
\begin{th}
The explicit form of the spin-shift operator for 
 the Ruijsenaars-Schneider Hamiltonian of type $C_n$ is
\begin{eqnarray*}
 D_{(l_{\alpha})}&=& \sum_{w \in W} {\rm sgn}(w)
  \Bigg(\,\prod_{\alpha \in \hat{R}_{w}^{-}}
   \frac{1- q^{2 l_{\alpha}} X^{\alpha}}{1-X^{\alpha}}
   \frac{1- q^{2(l_{\alpha}-1)} X^{\alpha}}{1-q^{-2} X^{\alpha}} \\
  && \times \prod_{s=1}^{2n-1} 
  \prod_{{\alpha \in \hat{R}_{w}^{-}} \atop {s+1 \le ht(-w^{-1}(\alpha))}}
  \frac{1- q^{2 (l_{\alpha}+s)} X^{-\alpha}}{1-q^{2 s}X^{-\alpha}}
   \frac{1- q^{2(l_{\alpha}-s-1)} X^{\alpha}}{1-q^{-2(s+1)} X^{\alpha}}
  \tau(w(\rho^{\vee})) \\
  && \quad + \sum_{\lambda^{\vee} \prec \rho^{\vee}} 
        G_{\lambda^{\vee}} \tau(\lambda^{\vee}) \Bigg).
\end{eqnarray*}
\end{th}

\makeatletter
 \def\@biblabel#1{#1.}

\makeatother

\end{document}